\documentclass[twocolumn,aps,prl,superscriptaddress]{revtex4-1}

\usepackage[english]{babel}
\usepackage[utf8]{inputenc}
\usepackage[T1]{fontenc}

\usepackage{hyperref}
\usepackage{amsmath, amssymb, amsfonts, mathrsfs}
\usepackage{xcolor,gensymb,epstopdf}
\usepackage{exscale,times}
\usepackage{graphicx,color}
\usepackage{latexsym}

\usepackage{dsfont}

\newcommand{\ket}[1]{|#1\rangle}

\begin{document}
\title{Experimental linear-optics simulation of the ground state of an interacting quantum spin-ring}

\author{Adeline Orieux}\email{adeline.orieux@gmail.com}
\affiliation{Dipartimento di Fisica, Sapienza Universit\`a di Roma, Piazzale Aldo Moro, 5, I-00185 Roma, Italy}
\author{Joelle Boutari}
\affiliation{Dipartimento di Fisica, Sapienza Universit\`a di Roma, Piazzale Aldo Moro, 5, I-00185 Roma, Italy}
\author{Marco Barbieri}
\affiliation{Clarendon Laboratory, University of Oxford, Parks Road, Oxford, OX1 3PU, United Kingdom}
\author{Mauro Paternostro}
\affiliation{Centre for Theoretical Atomic, Molecular and Optical Physics, School of Mathematics and Physics, Queen's University, Belfast BT7 1NN, United Kingdom}
\author{Paolo Mataloni}
\affiliation{Dipartimento di Fisica, Sapienza Universit\`a di Roma, Piazzale Aldo Moro, 5, I-00185 Roma, Italy}
\affiliation{Istituto Nazionale di Ottica, Consiglio Nazionale delle Ricerche (INO-CNR), Largo Enrico Fermi, 6, I-50125 Firenze, Italy}

\date{\today}

\begin{abstract}
Critical phenomena involve a structural change in the correlations of its constituents. These features can be employed in quantum simulators to assess when a criticality occurred in medium-size systems, for which phase transitions are not captured by standard thermodynamical methods. Here we demonstrate these concepts in a photonic quantum simulator: we encode the wavefunction of the ground state of a three-spin Ising ring by using a pair of entangled photons. The effect of a simulated magnetic field, leading to a critical modification of the correlation within the ring, is analysed by studying two- and three-spin entanglement. In particular, we make a quantitative connection between the violation of Svetlichny's inequality to the amount of three-tangle and tripartite negativity in our ring
\end{abstract}

\maketitle

{ Future quantum simulators will have the ability to investigate the statistical properties of strongly correlated medium-size systems for which explicit calculations are already impractical and mean-field approaches inadequate~\cite{Georgescu,Buluta,Walther}. Although in systems of only finite size the thermodynamical limit might be far from being achieved, interesting phenomena related to the appearance of long-range quantum mechanical correlations occur. Their reproduction in a quantum simulator will help us fully harnessing such critical manifestations typical of quantum many-body systems. 
}

In his seminal 1982 paper~\cite{Feynman}, Richard Feynman put forward the thought-provoking idea that more powerful synthetic emulations of quantum physics and chemistry would be possible by using quantum mechanical simulators rather than their classical versions. It took about twenty years before the vision inherent in such a suggestive proposal could see materialisation in the form of the experimental simulation (based on light-matter interactions) of simple yet very interesting quantum many-body effects~\cite{Friedenauer, Kim, Islam}. The nuclear magnetic resonance-based "digital" quantum simulation of thermal states of frustrated magnets has been reported~\cite{Zhang}. {The time evolution of spin chains up to six qubits has also been simulated in a trapped ion-based digital simulator~\cite{Lanyon2}.}

Photonics is one of the leading architectures for quantum simulators, due to the good level of control of the wavefunction and the high level of isolation from noise~\cite{AGW}. Demonstrations include the design and realisation of photonic settings for the quantum simulation of elementary quantum chemistry~\cite{Lanyon}, open-system dynamics~\cite{Chiuri}, and quantum random walks~\cite{Peruzzo, Sansoni, Crespi}, the latter exhibiting important Anderson-like disorder-induced effects. Frustration in a Heisenberg chain has been studied in an analog photonic quantum simulator~\cite{Ma}, while special topologically protected bound states appearing in condensed-matter physics have been emulated by the means of photonic quantum walks~\cite{Kitagawa}.

The reliable simulation of structural changes in the properties of the ground state of the system would allow to witness the emergence of critical manifestations typical of quantum many-body systems. In the proximity of the quantum critical point of a quantum spin model, in fact, long-range correlations of potentially quantum nature settle across the system as a result of the enhancement of quantum fluctuations associated with a quantum phase transition~\cite{Amico}. Usually, this is manifested in the observation of peculiar behaviours of figures of merit for the quantification of general quantum correlations (from entanglement to discord), as recently shown by a series of theoretical investigations~\cite{Rulli,Campbell1,Campbell2}.

While the majority of the studies reported in literature so far have dealt with the quantum correlations characterising the state of two spins picked from the many-body system, Refs.~\cite{Rulli,Paternostro,Campbell1,Campbell2} have shown that much information can be gathered from the study of global quantifiers that, by addressing the quantum correlation properties of the whole system, are able to faithfully characterise quantum criticality even in those situations where two-spin indicators fail, such as at high temperature. In this paper, we push forward this idea and, by implementing the analogue simulation of the ground state of a ferromagnetic Ising spin-chain, we experimentally characterise the structural changes that occur at the quantum phase transition from the ferromagnetic to the paramagnetic phase. 

\begin{figure*}[t!]
\includegraphics[width=\columnwidth]{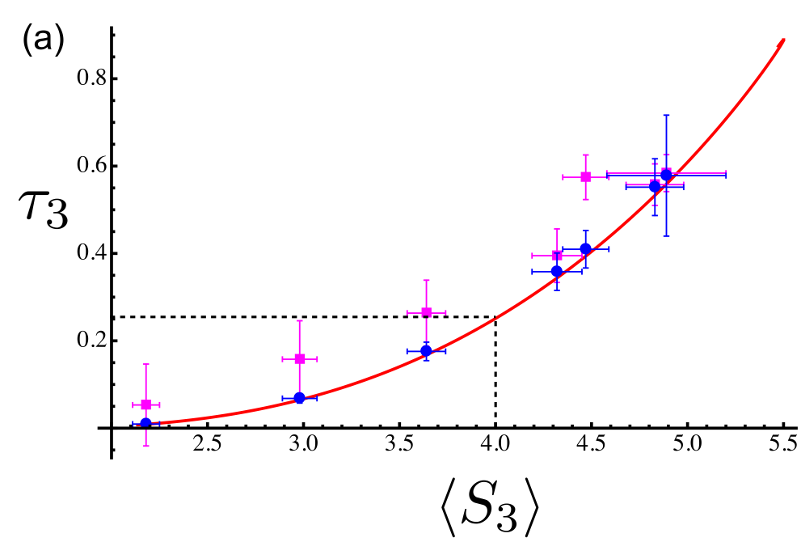}\includegraphics[width=\columnwidth]{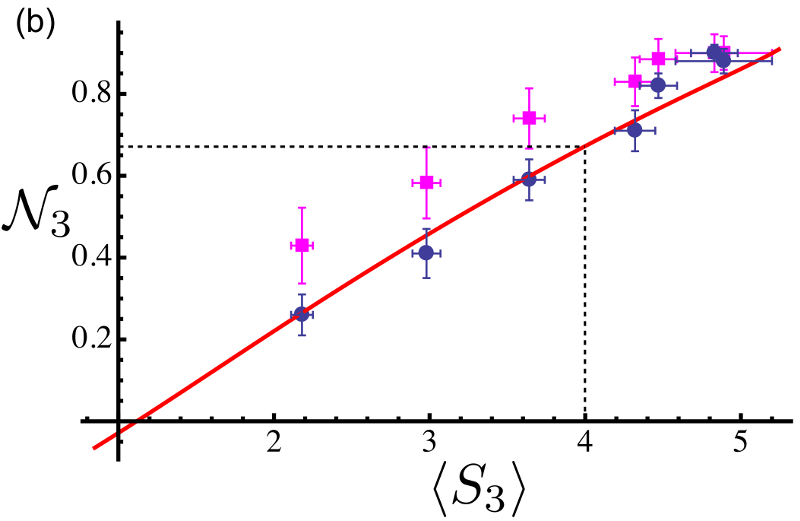}
\caption{{(a)} Relation between the Svetlichny parameter $|\langle S_3\rangle|$ and the three-tangle $\tau_3$ for the ground state $\ket{g_3(\beta)}$ of a three-spin Ising ring. The dashed vertical line marks the local realistic bound imposed to the Svetlichny inequality. In turn, this identifies the threshold value $\tau_3=0.25$ above which the three-spin state is non-local in a genuine tripartite sense. The (blue) circle-shaped points are the values of the three-tangle obtained using the analytic relation between $\tau_3$ and $|\langle S_3\rangle|$ discussed in the body of the paper, evaluated at  the experimental values of the Svetlichny parameter. The (magenta) square-shaped data points are the values of $\tau_3$ estimated using local measurement settings. (b) Analogous plot for the tripartite negativity ${\cal N}_3$. In this case, the threshold for tripartite non-locality is ${\cal N}_3\simeq0.67$. The same code of colors used in panel (a) holds here. Error bars are determined by standard error propagation with Poissonian distributions attached to the experimental counts.}
\label{3tangle}
\end{figure*}

Our analysis is conducted by experimentally assessing the multipartite Bell-like inequality embodied by the Svetlichny formulation~\cite{Svetlichny}. The violation of such inequality witnesses the occurrence, in the state of a system, of genuine multipartite non-local correlations. We show that the Svetlichny function evaluated using the ground state of the Ising chain violates the $N$-party local-realistic bound that, even for a short chain, is very close to the quantum critical point defined in the thermodynamic limit $N\to\infty$. Moreover, by exploiting the analytical link between the Svetlichny inequality and the measure of genuine multipartite entanglement embodied by the three-tangle~\ref{3tangle}, we estimate the degree of tripartite entanglement shared by the spins of an Ising chain at a set degree of violation of the Svetlichny inequality. Remarkably, such estimate is possible without a tomographically complete reconstruction of the ground state of the system. 

This investigation reveals the suitability of a linear-optics simulation of the collective features of a many-body system all the way across its quantum critical point and spurs tantalising extension of such analysis to the simulation of equilibrium states at non-zero temperature. In turn, this would enable the simulation of out-of-equilibrium thermodynamics in controllable table-top linear-optics set-ups by the means of recently proposed interferometric approaches to the reconstruction of the statistics following a quantum process~\cite{Dorner,Mazzola}.

\begin{figure}[t]
\centering
\includegraphics[width=0.7\columnwidth]{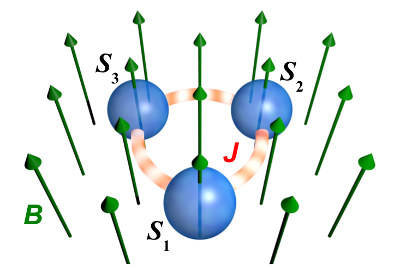}
\caption{{Three-spin Ising ring in a transverse magnetic field}. Schematic picture of the ring of three spins ($S_i$, $i=1,2,3$) under a transverse magnetic field $B$. The spin-spin coupling strength is labelled as $J$.}
\label{FigSpins}
\end{figure}

We have optically simulated the ground state of an Ising ring [cf. Fig. \ref{FigSpins}]
\begin{equation}
\label{Ising}
\hat{\cal H}_{I}=-{\cal J}\sum^N_{n=1}\hat\sigma^x_{n}\hat\sigma^x_{n+1}+{\cal B}\sum^N_{n=1}\hat\sigma^z_{n}
\end{equation}
with $\hat\sigma^k_i$ the $k$ Pauli operator of spin $i$ and $\hat\sigma^k_{N+1}\equiv\hat\sigma^k_{1}$~($k=x,y,z$). In Eq.~\eqref{Ising}, ${\cal J}$ is the inter-spin coupling strength and ${\cal B}$ is the magnetic energy of the spins subjected to a global transverse magnetic field. Besides being a key Hamiltonian in quantum statistical mechanics as it embodies one of the simplest models to show a phase transition, ${\cal H}_I$ has also attracted much attention from the quantum information community in light of the interesting quantum correlation properties of its ground state~\cite{Buzec,Amico}. In particular, for $\beta={\cal B}/{\cal J}\to 0$, the ground state $\ket{g_N(\beta)}$ of the Ising model approaches an $N$-spin state that is locally equivalent to the Greenberger-Horne-Zeilinger state
\begin{equation}
\ket{\text{GHZ}}=\frac{1}{\sqrt2}(\ket{+}^{\otimes N}+\ket{-}^{\otimes N})
\end{equation}
with $\ket{\pm}$ the eigenstates of $\hat\sigma^x$, thus exhibiting long range quantum correlations and multipartite quantum entanglement~\cite{Buzec}. While most of the attention has been focused on the behaviour of bipartite entanglement, it has been shown in Ref.~\cite{Paternostro} that $\ket{g(\beta)}$ is endowed with strong multipartite non-local properties, as witnessed by the violation of the Svetlichny inequality. For $\beta\to 0$, the degree of violation approaches the maximum allowed value of $4\sqrt{2^{N-k}}$ with $k=1$ ($k=2$) for an even (odd) number of spins, which is achieved for an $N$-spin GHZ state~\cite{Cereceda}, thus reinforcing the claim on the form of the ground state in such limit~\cite{Paternostro}. Quite remarkably, the violation of a generalised Svetlichny inequality and the onset of the features that are typical of a GHZ-like state occur close to $|\beta|=1$, which for a ring in the thermodynamic limit ($N\to\infty$) at zero temperature identifies the quantum critical point of the quantum Ising model at which a ferromagnetic-to-paramagnetic phase transition occurs~\cite{Sachdev}. For the finite-size chain addressed here (which, as we will see later on consists of only a few simulated spins), there cannot be a direct link between the establishment of such onsets and the model's quantum phase transition. A connection, on the other hand, should be searched with the changes of symmetries in the system occurring close to the critical point, which can in turn affect rather strongly the way quantum correlations are shared by the spins. 

\begin{figure*}[t!]
\centering
\includegraphics[width=16cm]{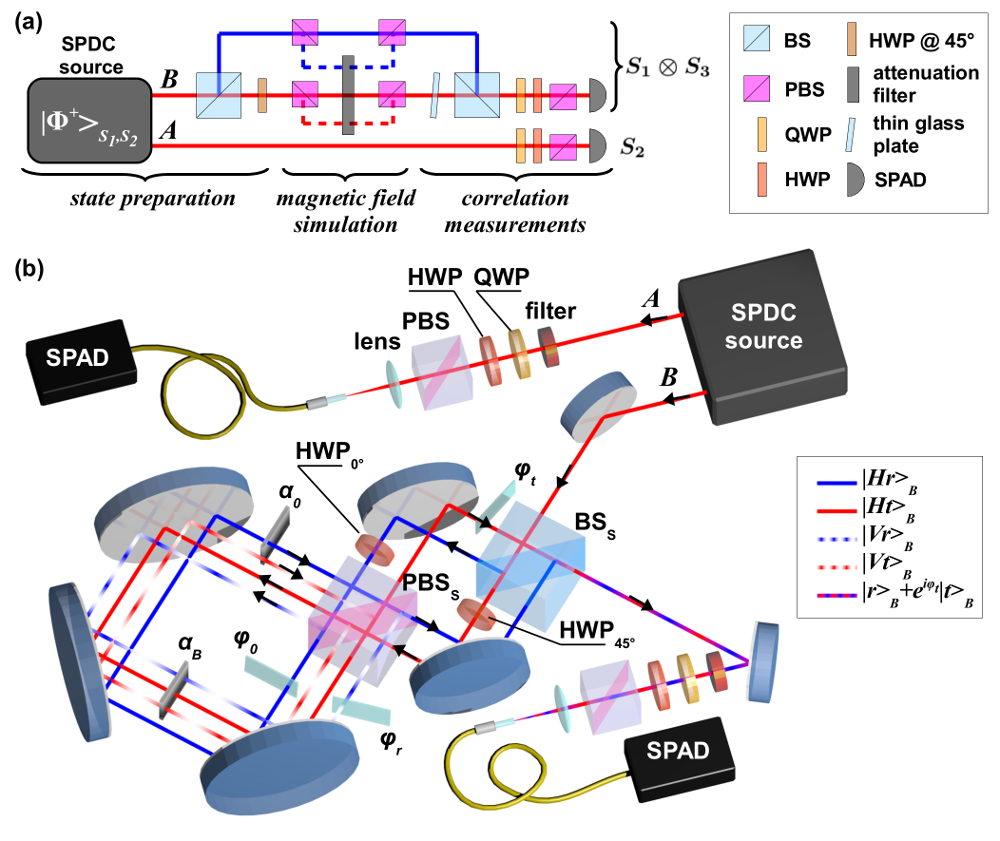}
\caption{{Photonic simulation of the ground state of a three-spin Ising ring in a transverse magnetic field}. (a) Conceptual set-up of the photonic simulation. A detailed description is given in the main text. SPDC source: spontaneous parametric down-conversion source generating photon pairs in the entangled state $\ket{\Phi^+}$ of spins $S_1$ and $S_2$. BS: beam-splitter; PBS: polarizing beam-splitter; QWP: quarter-wave plate; HWP: half-wave plate; SPAD: single photon avalanche photodiode. For photon B, the red (blue) line between the BSs corresponds to the transmitted (reflected) path (spin $S_3$), and the solid (dashed) line between the PBSs correspond to the horizontal (vertical) polarization of photon B (spin $S_1$). (b) Actual experimental implementation of the photonic simulator based on a doubly displaced Sagnac interferometer set-up. A detailed description is given in the Methods section. BS$_S$: beam-splitter; $\text{HWP}_{45^\circ}$ and $\text{HWP}_{0^\circ}$: $45^\circ$-oriented and $0^\circ$-oriented half-wave plates; PBS$_S$: polarizing beam-splitter; $\alpha_B$ and $\alpha_0$: attenuation filters; $\varphi_t$, $\varphi_0$ and $\varphi_r$: glass plates.}
\label{FigSet-up}
\end{figure*}

We have experimentally simulated the ground state of an $N{=}3$ spin Ising ring with ${\beta}<0$, which reads 
\begin{equation}
\ket{g_3(\beta)}=\frac{1}{\sqrt{3+a^2_0}}(a_0\ket{000}+\ket{011}+\ket{101}+\ket{110})_{123}
\end{equation}
with $a_0=-1-2\beta+2\sqrt{1+\beta+\beta^2}$. Our simulator makes use of a source of polarization-entangled photon pairs that generates by spontaneous parametric down-conversion the entangled state of spins 1 and 2 $|\Phi^+\rangle_{12}=\frac{1}{\sqrt{2}}(|00\rangle+|11\rangle)_{12}$ which is encoded in the polarization state of photons $A$ and $B$: $\frac{1}{\sqrt{2}}(|HH\rangle+|VV\rangle)_{BA}$ [cf. Fig.~\ref{FigSet-up} (a)]. Here, $|H\rangle$ and $|V\rangle$ design respectively the horizontal and vertical polarization states of photon A or B. The third spin is encoded in the path degree of freedom of photon B by having it go through a 50-50 beam splitter (BS), thus creating the state $\frac{1}{\sqrt{2}}(|r\rangle+|t\rangle)_{B}$, which encodes the logical state $\frac{1}{\sqrt{2}}(|0\rangle+|1\rangle)_{3}$. Finally, spin 3 is entangled to spin 1 by a controlled-NOT gate that is physically implemented between the path of photon B (the control spin) and its polarization (the target one): a half-wave plate (HWP) with its axis oriented at 45$^\circ$ with respect to the horizontal direction and inserted in the transmitted path of photon B performs a bit-flip on spin 1 depending on the state of spin 3. We thus obtain the state $|g_3(0)\rangle_{123}=\frac{1}{2}(|000\rangle+|110\rangle+|101\rangle+|011\rangle)_{123}\equiv\frac{1}{2}(|HHr\rangle+|VVr\rangle+|VHt\rangle+|HVt\rangle)_{BAB}$ which corresponds to the ground state of the spin ring with no magnetic field and is locally equivalent to $\ket{\text{GHZ}}_{123}$ upon application of the three-spin Hadamard transform $[(\sigma^x+\sigma^z)/\sqrt{2}]^{\otimes3}$. 

The transverse magnetic field, resulting in non-zero values of $\beta$ and the possibility to explore the properties of the ground state of the Ising ring, is then simulated by introducing a variable attenuation on photon B depending on the joint state of spins 1 and 3. This is achieved by first splitting the four path-polarization components of photon B with two polarizing beam-splitters (PBS) and then inserting a  variable attenuation filter on the modes corresponding to $|011\rangle_{123}$, $|110\rangle_{123}$ and $|101\rangle_{123}$. Correlations between the three spins can then be measured through coincidence measurements in different projection basis

This conceptual scheme is implemented in a doubly displaced Sagnac interferometer [cf. Fig.~\ref{FigSet-up} (b)] to guarantee perfect phase stability of the state, and is described in the Appendix. The tunability of the parameters that enter our experimental simulator allowed us to achieve $\beta\in[-2,0]$, which are sufficient to explore the most salient features of the model at hand.

\begin{figure*}[t!]
\centering
\includegraphics[width=16cm]{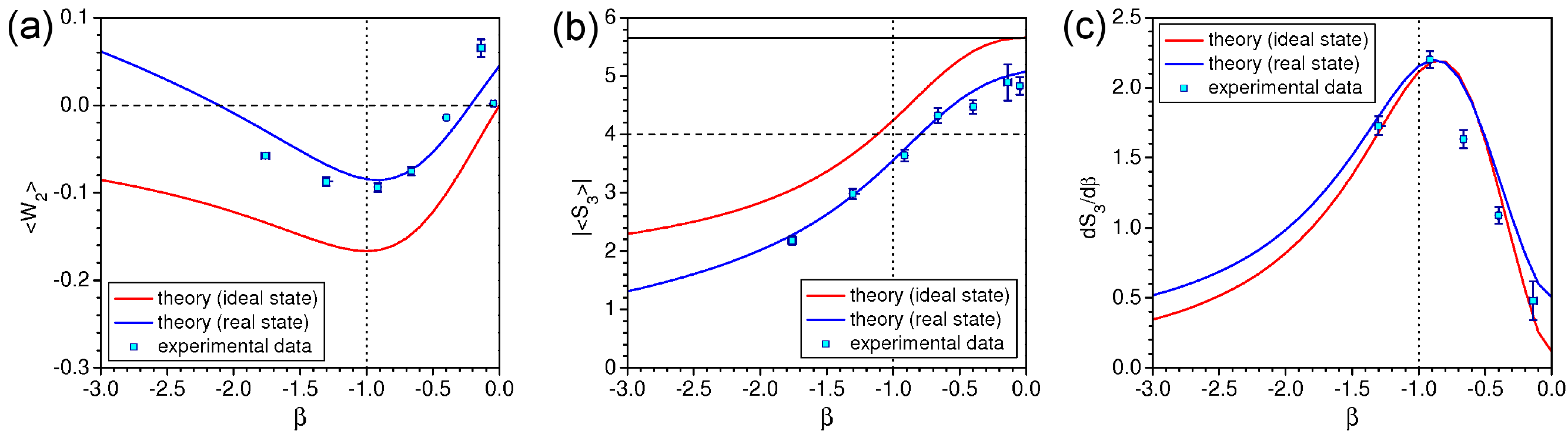}
\caption{{Measurement results.} Measured evolution of the expectation value of (a) the bipartite entanglement witness $\langle W_2\rangle$, (b) the Svetlichny function $|\langle S_3\rangle|$ and (c) its derivative $\partial_\beta{|\langle S_3\rangle|}$ as a function of $\beta$. The square dots correspond to measurements performed on the optically simulated ground state; the red line shows the theoretically expected evolution for the ideal ground state; and the blue line represents the theoretical evolution for a state affected by noise, as is the case in the experiment. The vertical dotted line correspond to the critical point $\beta=-1$; the horizontal dashed line in (a) and (b) show the lowest (highest) possible value that can be reach by $\langle W_2\rangle$ ($|\langle S_3\rangle|$) for a (bi-)separable state; and the full black horizontal line in (b) shows the highest value that can be reached by a GHZ state.}
\label{FigResultats}
\end{figure*}

Our goal is to characterise the fundamental symmetry changes occurring in the ground state of the Ising ring when crossing the quantum critical point by assessing multipartite non locality and entanglement. In order to achieve this, we have measured the amount of both bipartite and tripartite entanglement in the simulated $\ket{g_3(\beta)}$ for each value of $\beta$ in our experiment. We perform our characterisation of the quantum correlation properties of the model in a gradual manner, starting for an assessment of the bipartite entanglement between any two spins taken out of the ring. This is pursued by considering the entanglement witness operator 
$\hat W_{2}=|v_{2}\rangle\langle v_{2}|^{\text{\sf pt}}$, where $|v_{2}\rangle$ is the eigenvector of $\rho^{\text{\sf pt}}_2(\beta)$ associated with the smallest eigenvalue~\cite{Guhne_witness}. Here, {\sf pt} stands for the partial transposition operation and $\rho_2(\beta)$ is the two-spin reduced state 
\begin{equation}
\rho_{ij}(\beta)=\frac{1}{a_0^2+3}
\begin{pmatrix}
a_0^2 & 0 & 0 & a_0\\
 0 & 1 & 1 & 0\\
 0 & 1 & 1 & 0\\
  a_0 & 0 & 0 & 1
\end{pmatrix},~~\text{with}~~i\neq j=1,2,3.
\end{equation}
Achieving $\langle\hat W_2\rangle<0$ guaranties bipartite entanglement in the state of the spin pair. Although the translational invariance of the Ising ring makes any choice of spin pairs equivalent, in our experiment we have considered spins 2 and 3 and decomposed the witness into a combination of local measurement settings as 
\begin{equation}
\langle \hat W_{2} \rangle=\frac{1}{4}\langle \left(\mathds{\hat 1}_2\otimes\mathds{\hat 1}_3 - \hat \sigma_2^x\otimes\hat \sigma_3^x + \hat \sigma_2^y\otimes\hat \sigma_3^y - \hat \sigma_2^z\otimes\hat \sigma_3^z \right)\rangle.
\end{equation}
When calculated over separable states, this expectation value is always positive. Therefore, a negative value signals the existence of a non-zero amount of bipartite entanglement.

As anticipated, the focus of our simulation is to address critical structural changes on the multipartite non-local properties of the simulated ground state. This is done by demonstrating experimentally the violation of the tripartite Svetlichny inequality~\cite{Svetlichny}, which would witness genuine tripartite non-locality in the same manner as the Bell inequality does for bipartite non-locality. The inequality is written as $|\langle{S}_3\rangle|=|\langle M_3+M'_3\rangle|\leq4$ with the Svetlichny function $\langle S_3\rangle$ that can be built as the combination of the Mermin-Ardehali-Belinskii-Klyshko functions~\cite{Mermin,Klyshko} $\langle M_3\rangle$ and $\langle M'_3\rangle$ given in the Appendix~\cite{Cereceda}. The Svetlichny inequality has been violated using a photonics GHZ state in Ref.~\cite{Lavoie}. 
 
 For the ground state of a three-spin ring, it is straightforward to show that  
\begin{equation}
\begin{aligned}
\langle\hat S_3\rangle=&\sqrt{2}\left(\langle\hat\sigma_1^y\otimes\hat\sigma_2^z\otimes\hat\sigma_3^y\rangle +\langle\hat\sigma_1^z\otimes\hat\sigma_2^y\otimes\hat\sigma_3^y\rangle\right.\\
&\left. + \langle\hat\sigma_1^y\otimes\hat\sigma_2^y\otimes\hat\sigma_3^z\rangle - \langle\hat\sigma_1^z\otimes\hat\sigma_2^z\otimes\hat\sigma_3^z\rangle\right),
\end{aligned}
\end{equation}
which can be easily measured in our set-up by implementing four local measurement settings. From the knowledge of $\langle\hat S_3\rangle$ it is also possible to estimate the value of measures of genuine tripartite entanglement. For instance, for the specific case of a three-spin ring, we can link the violation of the Svetlichny inequality to the measure of tripartite entanglement embodied by the {\it three-tangle} $\tau_3={\cal C}^2_{1|23}-\sum^3_{j=2}{\cal C}^2_{1|j}$~\cite{Coffman} with ${\cal C}_{1|j}$ the concurrence of the bipartite state composed of spin 1 and $j=2,3$ and ${\cal C}_{1|23}=2\sqrt{\det\rho_{1}}$ that measures the entanglement between spin $1$ and the two-spin system composed of particles $2$ and $3$. A second measure that can be linked to $\langle S_3\rangle$ is the {\it tripartite negativity} ${\cal N}_3=\sqrt[3]{{\cal N}_{1|23}{\cal N}_{2|13}{\cal N}_{3|12}}$ with ${\cal N}_{i|jk}$ the negativity of the bipartite system composed of spin $i$ and $(j,k)$~\cite{Alcaine}.

It is straightforward to relate analytically the degree of genuine tripartite entanglement as quantified by $\tau_3$ and ${\cal N}_3$ to the values taken by the Svetlichny function $\langle{S}_3\rangle$ as 
\begin{equation}
\label{formule}
\begin{aligned}
&\tau_3
=\frac{3 ({|\langle{S}_3\rangle|}^2-2 \sqrt{2}|\langle{S}_3\rangle|-4)+\sqrt{3|\langle{S}_3\rangle|} \left(4 \sqrt{2}-|\langle{S}_3\rangle|\right)^{\frac32}}{36},\\
&{\cal N}_3=\frac{2\sqrt{2(1+a^2_0)}}{3+a^2_0}\\
&\text{with}~~a_0=\frac{3\sqrt2+\sqrt{12\sqrt2 |\langle S_3\rangle|-3|\langle S_3\rangle|^2}}{|\langle S_3\rangle|-\sqrt2}.
\end{aligned}
\end{equation}
The functional link between such measures of multipartite entanglement and the Svetlichny function is shown in Fig.~\ref{3tangle}, where we have identified a threshold value of $\tau_3$ and ${\cal N}_3$ above which genuine tripartite non locality is ensured. Notice that, while the value of $\tau_3$ is strongly determined by the similarity between our resource state and a GHZ state (and is null, for instance, for a $W$ state such as $(\ket{011}+\ket{101}+\ket{110})/\sqrt3$), this is not the case for ${\cal N}_3$, which can attain significant values also for other forms of multipartite entangled states, not necessarily of the GHZ form.

The results of the measurements performed on the simulated state are presented in Fig.~\ref{FigResultats} as a function of $\beta$, together with the theoretical predictions. Moreover, we have been able to provide an estimate of the values taken by the three-tangle of the state of the simulated chain at hand made on the basis of the experimentally measured values of $\langle S_3\rangle$ together with a comparison with the value of $\tau_3$ that could be reconstructed putting together the outcomes of our experimental measurements of multi-spin correlation functions [cf. Fig.~\ref{3tangle} {\bf (a)}]. 

However, an important point  should be remarked: a fair comparison between experimental results and theoretical prediction will have to take into account the imperfections of the optically simulated ground state with respect to the ideal one $|g(\beta)\rangle$. In particular, the real photonic state is affected by background noise that mixes the ground state of the spin ring to white noise, so to obtain the Werner state $\rho_{exp}(\beta)=p|g(\beta)\rangle\langle g(\beta)|+(1-p)\mathds{1}/8$, where $p={R_{SNR}}/({2+R_{SNR}})$ depends on the signal to noise ratio $R_{SNR}$. In our simulator, the value of the effective magnetic field is changed by attenuating the signal in specific paths in the interferometers. This implies that there is less signal in those paths for large values of $|\beta|$ than for small ones, which results in a $R_{SNR}$ that diminishes with increasing values of $|\beta|$. In turn, this makes $p$ dependent on $\beta$. We have experimentally found that $p=0.128\beta+0.927$, which allowed us to compute the blue theoretical curves in the graphs. The good agreement between the measured and theoretical values confirms that our photonic simulator does mimic well the targeted three-spin ring ground state. 

As it can be seen from the graphs, the quantum correlations of the ground state of the ring undergo a profound modification at the critical point. First, the bipartite non-locality [Fig. \ref{FigResultats} (a)], which is null both for $\beta\to0$ and $\beta\rightarrow-\infty$, reaches a non-null value close to the point of structural changes for the model which would correspond to the critical point at the thermodynamic limit. At the same time, the tripartite non-locality witnessed by the Svetlichny function [Fig. \ref{FigResultats} (b)], which is maximum for $\beta\to0$ when the ground state is close to a GHZ state, decreases continuously as $|\beta|$ increases and reduces to values lower than the non-locality threshold just after the structural-change point. The inflexion point that is visible in the plot of $|\langle S_3\rangle|$ against $\beta$, close to the expected critical point, reveals in turn that the derivative $\partial_\beta{|\langle S_3\rangle|}$ will have a maximum at that point. This is clearly seen in Fig.~\ref{FigResultats} (c), which demonstrates the sensitivity of the figure of merit embodied by $\partial_\beta{|\langle S_3\rangle|}$ to the  modifications undergone by the structure of multipartite quantum correlations of our system in the proximity of the `critical point' $\beta\simeq-1$. 

Let us also notice that the noise affecting the simulated state only induces a quantitative reduction on the measured Svetlichny function but does not change the value of $\beta$ at which the criticality-induced modifications to the sharing of quantum correlations occur. The maximum of $|\langle W_2\rangle|$ and the point of inflexion of $\langle S_3\rangle$ are found at $\beta\simeq-1$ independently of the amount of noise affecting the state. This guarantees the correct detection of the critical point even though the system is not perfectly simulated, which is encouraging for potential medium-size system simulations for which no classical computation can be made. On the other hand, the mixed nature of the state affected by noise demands the modification of the definition of $\tau_3$ and its reformulation in terms of convex-roof extensions~\cite{Coffman}. The calculation of three-tangle for general three-qubit states is in general a very demanding task that goes beyond the scopes of our experimental work. However, the analysis of the results presented in Fig.~\ref{3tangle} (a) shows that both our experimental reconstruction and the values estimated using Eq.~\eqref{formule} are very close to the behavior expected for the proper ground state of the simulated chain, hinting strongly at the very good quality of the experimental data.  Moreover, ${\cal N}_3$ is well suited for mixed states of three spins without modifications to the definition provided above. This has enabled us to generalise the link between tripartite negativity and Svetlichny function so to provide a non-tomographic estimate of the amount of tripartite entanglement content of $\rho_{exp}(\beta)$ based on the experimentally measured values of $\langle S_3\rangle$ [cf. Table~\ref{default}]. These values are reported in Fig.~\ref{3tangle} (b), showing the excellent agreement of our estimates and the expected functional relation between ${\cal N}_3$ and the Svetlichny parameter. 

We have experimentally studied multipartite non-locality in a short Ising ring, whose ground state undergoes important changes, at the level of the quantum-correlation sharing, as the model approaches its critical point. Our study has been enabled by the experimental implementation of a flexible photonics quantum simulator that, by exploiting the potential of reliable hyper-entangled resources, is able to reproduce the salient features of the ground state of the Ising model in an ample range of experimental values. We believe that, when extended to the simulation of equilibrium states at a non-zero effective temperature, our study paves the way to the experimental investigation of interesting many-body effects in photonics quantum simulators~\cite{Walther}, from criticality to quantum thermodynamics.

\section*{Appendix}

{\bf Experimental set-up details}.---The experimental photonic quantum simulator consists in the doubly displaced Sagnac interferometer set-up presented in Fig.~\ref{FigSet-up} (b) that allows to prepare the two-photon three-qubit state $|g_3(0)\rangle_{123}$, simulate a transverse magnetic field and measure two-photon coincidence detection events in different projection basis of the three spins. The conceptual set-up was already described in the main text and in Fig.~\ref{FigSet-up} (a); here we explicit the correspondence between the actual Sagnac-based set-up and the former and give details on the role played by the different optical elements.

The preparation of $|g_3(0)\rangle_{123}$ is achieved thanks to the beam-splitter $\textrm{BS}_\textrm{S}$ and the 45$^\circ$-oriented half-wave plate ($\textrm{HWP}_{45^\circ}$) in the transmitted path. The optical delay introduced by this HWP was compensated by a second HWP with its axis at 0$^\circ$ with respect to the horizontal direction in the reflected path ($\textrm{HWP}_{0^\circ}$).

The transverse magnetic field is then simulated by splitting the polarization components of photon B, both for the reflected and transmitted path, with the polarizing beam splitter $\textrm{PBS}_\textrm{S}$ and introducing a variable attenuation filter ($\alpha_B$) common to the paths $|Ht\rangle_{B}$, $|Vr\rangle_{B}$ and $|Vt\rangle_{B}$. Again, a second glass slab ($\alpha_0$) on the path $|Hr\rangle_{B}$ is used to maintain paths of equal optical length in the interferometer. The polarization components are then recombined on the PBS before projective measurements are performed.

Let us notice that, because of an unbalance in the BS double transmission ($\eta_{t}/2=66\%$) and double reflection ($\eta_{r}/2=34\%$) coefficients, as well as in the SPDC source HH-cone ($\eta_{HH}/2=58\%$) and VV-cone ($\eta_{VV}/2=42\%$) emission probabilities, the state that is actually generated in our experiment is 
\begin{equation}
\begin{aligned}
|g_{3exp}(0)\rangle_{123}&={\cal N}\left(\sqrt{\eta_{HH}\eta_{r}}a_0|000\rangle+\sqrt{\eta_{VV}\eta_{r}}|110\rangle\right.\\
&\left.+\sqrt{\eta_{HH}\eta_{t}}|101\rangle+\sqrt{\eta_{VV}\eta_{t}}|011\rangle\right)_{123}.
\end{aligned}
\end{equation}
with ${\cal N}^{-1}=\sqrt{\eta_{VV}\eta_{r}+\eta_{HH}\eta_{t}+\eta_{VV}\eta_{t}+\eta_{HH}\eta_{r}a_0^2}$. This is why we chose to trace out qubit 1 for the evaluation of the bipartite entanglement witness $W_{2}$: we can see that in this case we obtain more balance between the $\ket{\Phi^+}$ and $\ket{\Psi^+}$ parts of the reduced density matrix $\rho_2$, as is the case in the ideal ground state.
The correspondence between the attenuation coefficient  and the simulated value of $a_0$ (and thus of $\beta$) is given by $\alpha{=}\frac{P_{011}+P_{101}+P_{110}}{P_{000}} {=}\frac{\eta_{VV}\eta_{t}+\eta_{HH}\eta_{t}+\eta_{VV}\eta_{r}}{\eta_{HH}\eta_{r}a_0^2}$, where $P_{ijk}$ stand for the probability of two-photon coincidence of state $|ijk\rangle_{1,2,3}$.
In the experiment, we could vary $\alpha$ from 0.1 to 4, and thus achieve values of $\beta$ within the interval $[-2;0]$.

Finally, projective measurements are done on both photons before registering their coincidence detection with two single photon avalanche photodiodes (SPAD) and a coincidence counting electronics. For both polarization qubits, this is achieved by a standard polarization analysis set-up consisting of a quarter-wave plate (QWP), a HWP and a PBS, while for the path qubit the projection on different basis is achieved by the second passage through $\textrm{BS}_\textrm{S}$ and a glass plate ($\varphi_t$) that can be rotated in the transmitted path so as to change the relative phase between $\ket{r}_B$ and $\ket{t}_B$. The optical delay introduced by this glass plate is compensated by two glass plates ($\varphi_0$ and $\varphi_r$) inserted in the reflected path. Note that these two glass plates are also used to correct for the $\pi$ phase difference introduced by $\textrm{HWP}_{0^\circ}$ between the horizontal and vertical polarization components of the reflected path.

\begin{table}[h!]
\begin{center}
\begin{tabular}{c|c} \hline\hline
$\langle S_3\rangle_{exp}$&${\cal N}_3$\\ \hline
4.83$\pm$0.15&0.90$\pm0.02$\\
4.89$\pm$0.31&0.88$\pm0.03$\\
4.47$\pm$0.12&0.82$\pm0.03$\\
4.32$\pm$0.13&0.71$\pm0.05$\\
3.64$\pm$0.10&0.59$\pm0.05$\\
2.98$\pm$0.09&0.41$\pm0.06$\\
2.18$\pm$0.07&0.26$\pm0.05$\\
 \hline\hline
\end{tabular}
\end{center}
\caption{Table for the estimates of the tripartite negativity in the noise-affected ground state of a three-spin Ising ring corresponding to the experimental values of the Svetlichny function, here dubbed $\langle S_3\rangle_{exp}$, measured at the values of $\beta$ reported in Fig.~\ref{FigResultats}.}
\label{default}
\end{table}

{\bf Expression of the Svetlichny inequality}.---The expression of the Svetlichny inequality for the three-spin ring is given by ${\cal S}_3=|M_3+M'_3|$ with $M_3=E(a_1,b_1,c_2)+E(a_1,b_2,c_1)+E(a_2,b_1,c_1)-E(a_2,b_2,c_2)$ and $M'_3=E(a_2,b_2,c_1)+E(a_2,b_1,c_2)+E(a_1,b_2,c_2)-E(a_1,b_1,c_1)$~\cite{Svetlichny,Cereceda,Paternostro} the Mermin-Ardehali-Belisnskii-Klyshko function~\cite{Mermin,Klyshko}. Here, $E(a_i,b_j,c_k)$ ($i,j,k=1,2$) is the statistical correlation function for local spin measurements having outcomes $a_i$, $b_j$, and $c_k$ respectively. In the case of the ground state of the N=3 Ising ring, it can be shown that ${\cal S}_3=|M_3+M'_3|$ is maximal for $a_1=3\pi/4$, $a_2=\pi/4$, $b_1=c_1=\pi/4$ and $b_2=c_2=-\pi/4$. With these values, $E(a_i,b_j,c_k)$ ($i,j,k=1,2$) can be expressed with the Pauli matrices $\hat\sigma^y$ and $\hat\sigma^z$: for example we have $E(a_1,b_1,c_1)=\frac{1}{2\sqrt{2}}\langle(\hat\sigma^y_1-\hat\sigma^z_1)\otimes(\hat\sigma^y_2+\hat\sigma^z_2)\otimes(\hat\sigma^y_3+\hat\sigma^z_3)\rangle$.


\section*{Acknowledgements}

MB is grateful to the Quantum Optics group at Sapienza, Universita' di Roma for hospitality during the development and completion of this work, and acknowledges support by the UK EPSRC (EP/H03031X/1 and EP/K034480/1). MP thanks the UK EPSRC for a Career Acceleration Fellowship and a grant awarded under the ``New Directions for Research Leaders" initiative (EP/G004579/1), the Alexander von Humboldt Stiftung, and the John Templeton Foundation (grant 43467). This work was supported by the EU-Project CHISTERA-QUASAR, PRIN 2009 and FIRB-Futuro in ricerca HYTEQ.

\end{document}